\documentclass[twocolumn]{aastex63}
\usepackage{amsmath}

\newcommand{\tbf}{}
\pdfoutput=1

\graphicspath{{./}{figures/}}

\accepted{for publication in AJ, 11/06/19}

\shorttitle{Long Period Planets in Overlapping Fields}
\shortauthors{S. Dholakia et al.}

\begin{document}

\title{Constraining Orbital Periods from Nonconsecutive Observations: \\ Period Estimates for Long-Period Planets in Six Systems Observed by K2 During Multiple Campaigns}

\correspondingauthor{Shishir Dholakia}
\email{dholakia.shishir@berkeley.edu}

\author{S. Dholakia} \altaffiliation{These authors contributed equally to this work}
\affiliation{Department of Astronomy, University of California, Berkeley 94720}

\author{S. Dholakia} \altaffiliation{These authors contributed equally to this work}
\affiliation{Department of Astronomy, University of California, Berkeley 94720}

\author{Andrew W. Mayo}
\affiliation{Department of Astronomy, University of California, Berkeley 94720}

\author{Courtney D. Dressing}
\affiliation{Department of Astronomy, University of California, Berkeley 94720}





\begin{abstract}
Most planetary discoveries with the K2 and {\em TESS} missions are restricted to short periods because of the limited duration of observation. However, the re-observation of sky area allows for the detection of longer period planets. We describe new transits detected in six candidate planetary systems which were observed by multiple K2 mission campaigns.  One of these systems is a multiplanet system with four candidate planets; we present new period constraints for two planets in this system. In the other five systems, only one transit is observed in each campaign, and we derive period constraints from this new data. The period distributions are highly multimodal resulting from missed potential transits in the gap between observations. Each peak in the distribution corresponds \tbf{to transits at} an integer harmonic of the two observed transits. We further detail a generalized procedure to constrain the period for planets with multiple observations with intervening gaps. Because long period photometrically discovered planets are rare, these systems \tbf{are interesting targets for follow-up observations} and confirmation. \tbf{Specifically, all six systems are bright enough (V = 10.4-12.7) to be amenable to radial velocity follow-up.} This work serves as a template for period constraints in a host of similar yet-to-be-discovered planets in long baseline, temporally gapped observations conducted by the \em{TESS} mission.
\end{abstract}
\section{Introduction} \label{sec:intro}
Over the span of 3.5 years and a data analysis effort extending until the present, the {\em Kepler} Mission has discovered \tbf{2345}\footnote{\label{nea}https://exoplanetarchive.ipac.caltech.edu as of \tbf{25 October} 2019} exoplanets. Due to the failure of two of the reaction wheels on {\em Kepler}, the K2 mission \tbf{was} devised to continue observations using the spacecraft thruster to correct for spacecraft roll \citep{howell2014}. K2 was designed to observe regions of the sky near to the ecliptic for approximately 75 days before moving on to a new location. 

Over a fifth of the planet candidates discovered by the original {\em Kepler} Mission had periods longer than 100 days, owing to the 4 year baseline of observations$^{\ref{nea}}$. However, as a result of the shorter baseline, the K2 mission was focused on finding planets with a period \tbf{$<$ 30 days}, in order to observe multiple transits in a baseline.

However, several K2 campaigns were overlapped partially or fully to cover certain high interest regions, including the Hyades and Praesepe open clusters. This provides a useful test of space-based data consisting of baselines separated by long gaps, which \tbf{is a} very similar method of observation \tbf{to} the {\em TESS} mission. For instance, 6167 targets in Campaigns 5 and 16 of the {\em K2} mission were observed in both campaigns. Furthermore, Campaign 5 and 18 had almost full overlap in their mission fields.
\begin{figure}[h]
    \centering
    \plotone{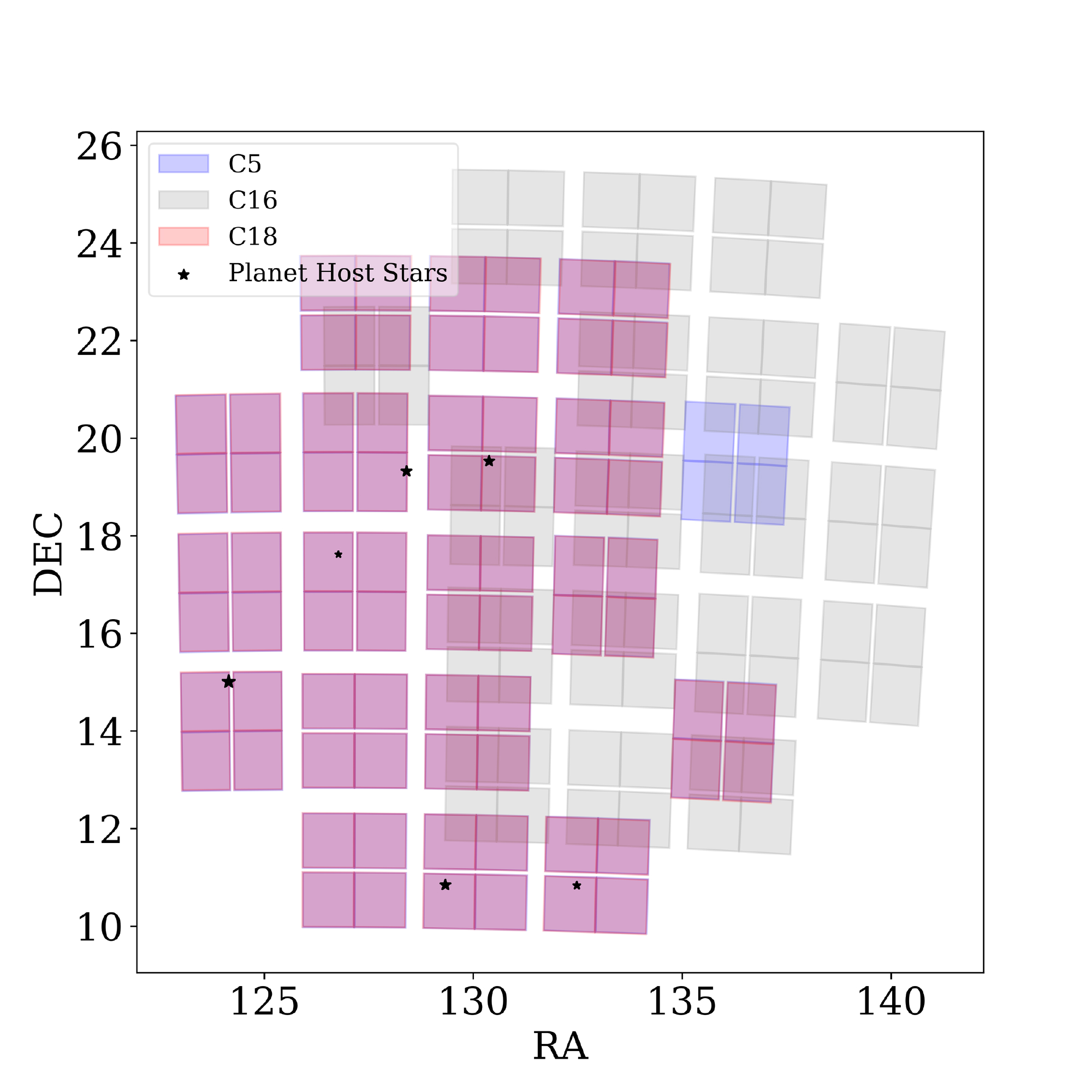}
    \caption{On-sky overlap of C5, C16, and C18. C5 and C18 overlap nearly completely. Note: Star sizes correspond to the Gaia G magnitude.}
    \label{fig:overlapplot}
\end{figure}

The {\em TESS} mission has a similar scope and design to the K2 mission, observing each sector for 27 days. Due to the {\em TESS} scanning strategy, the ecliptic poles will be observed nearly continuously, whereas regions close to the ecliptic plane will be observed for only a single sector. As a result, the focus of {\em TESS} in regions close to the ecliptic will be to find planets with periods $<$ 10 days, some of which may be in the habitable zones of cooler M-type stars \citep{tess2014}.

\tbf{In this paper we search K2 campaigns with overlapping fields for systems with single transits in more than one campaign. After identifying these systems, we characterize the planet candidates, including constraining their periods via an analytical period prior and transit modeling.}

\begin{figure*}[h]
    \centering
    \plotone{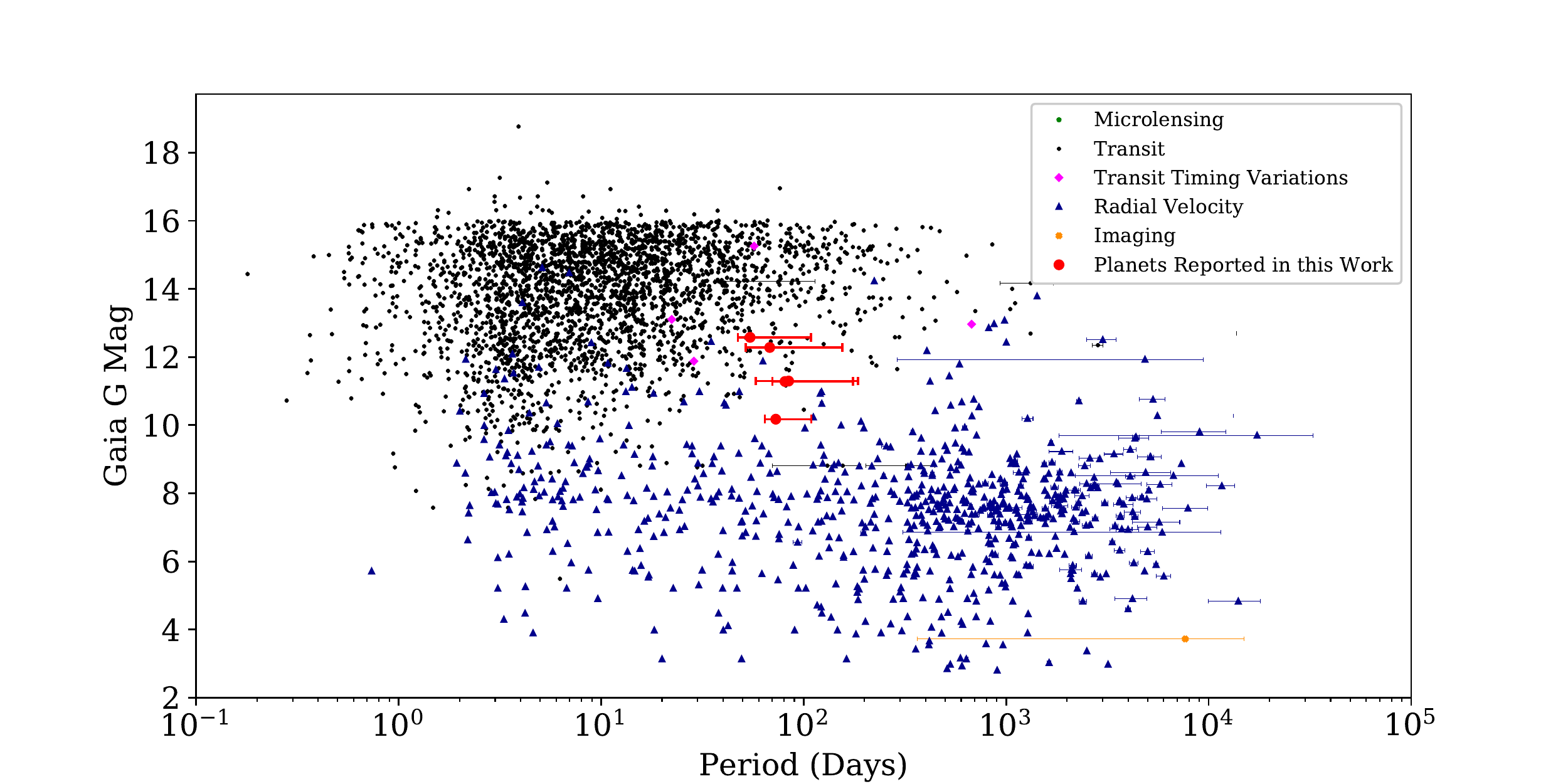}
    \caption{Plot showing parameter space for brightness and period for all planets found using common discovery methods as of Feb 2018 (https://exoplanetarchive.ipac.caltech.edu). The red bars represent planet candidates in this work; the bars occupy the 95\% confidence interval for period.}
    \label{fig:period_paramspace}
\end{figure*}

\tbf{We describe the search of K2 campaigns in Section \ref{sec:photometry}. We identify six such systems, presented in Section \ref{sec:planetcandidates}. In Section \ref{sec:periodconstraints} we present a generalized period prior that quantifies the probability of a period given that transits may have been missed in the interval between observations. We apply this period prior to obtain posterior period distributions for all six systems in Figure \ref{fig:periods}. In Section \ref{sec:summary} we discuss followup potential and the usefulness of the methodology of this paper for the TESS mission. The systems we present also fulfill an important region of parameter space in the search for exoplanets that are good targets for detailed characterization, shown in Figure \ref{fig:period_paramspace}.}

\section{K2 Photometry} \label{sec:photometry}

We performed a search of overlapping targets in \tbf{Campaign 5 (C5), Campaign 16 (C16)}, and Campaign 18 (C18), aiming to find longer period planets which may have \tbf{1 or 2 transits} in each campaign. We focused specifically on stars for which \citet{lacourse2018} detected a single transit event in order to determine whether additional transits were observed during C16, C18, or both campaigns.

We created lightcurves using the Python package \tbf{\texttt{lightkurve}} \citep{lightkurve} and also used lightcurves from both the K2SC pipeline \citep{Aigrain2016} as well as the K2SFF pipeline of \citet{vanderburg2014}. For most of the stars studied, we used the K2SC pipeline to generate lightcurves for both campaigns. \tbf{In some cases, the pipeline mask for lightcurves in Campaign 5 were suboptimal. As such, the K2SC detrended lightcurves, which start with the raw pipeline lightcurves, displayed poorer precision. For these cases, we used the K2SFF lightcurves instead, which use a custom mask optimized for precision \citep{vanderburg2014}.} For K2SFF lightcurves, we used an iterative version of the Savitsky-Golay filter \citep{savitzkygolay1964} implemented in the \texttt{lightkurve} pipeline in order to detrend the lightcurve. We first removed a trend and used the initial trend removal to mask any points farther than 4 sigma from the median of the lightcurve. Then, we performed another trend removal with the outlier points (including transits) masked. Finally, we interpolated and divided out the trend.

\section{Planet \tbf{Candidates} and Double Transiting Events} \label{sec:planetcandidates}

We searched \tbf{by eye} 24 \tbf{light curves with} single transit events reported by \citet{lacourse2018} in C5 data in order to check for additional transits. We found 6 systems which present additional transit events in either C16, C18, or both, which are plausibly planetary in nature. Five of these systems exhibit only a single additional transit in a subsequent campaign, but one of them, EPIC211939692, exhibits multiple additional transits in C18. EPIC211953574 was observed for C5, C16 and C18; the remaining five targets were observed in C5 and C18 alone. \tbf{Stellar parameters are derived in Section ~\ref{sec:stellarparams} and are presented in Table ~\ref{tab:tab1}.}

\begin{deluxetable*}{ccccccccc}
\tablecolumns{9}
\tablewidth{0pt}
\tablecaption{Stellar Parameters}
\tablehead{
\colhead{EPIC ID} & \colhead{RA} & \colhead{DEC} & \colhead{$R_{*}$/$R_\odot$} & \colhead{\tbf{Gaia} G mag}& \colhead{\tbf{Gaia T$_{eff}$}} & \colhead{\tbf{Gaia Luminosity} ($L_\odot$)} & \colhead{\tbf{$M_{*}$/$M_\odot$}$^1$} & \colhead{\tbf{$R_{*}$/$R_\odot$ (V-K)}$^1$}
}
\startdata
  211939692 & 128.401884 & 19.320186 & 1.29$^{+0.11}_{-0.20}$ & $11.718\pm0.0003$ & 6417$^{+120}_{-72}$ & $2.56\pm0.07$ & $1.22\pm0.02$ & $1.38\pm0.10$ \\
  211953574 & 130.380628 & 19.526300 & 1.11$^{+0.03}_{-0.04}$ & $11.2943\pm0.0007$ & 5924$^{+630}_{-280}$ & $1.37\pm0.02$ & $1.07\pm0.005$ & $1.11\pm0.04$ \\
  211821192 & 126.773816 & 17.618253 & 0.98$^{+0.03}_{-0.01}$ & $12.5766\pm0.0002$ & 5786$^{+20}_{-90}$ & $0.97\pm0.02$ & $0.99\pm0.005$ & $1.04\pm0.06$ \\
  211633458 & 124.146077 & 15.005897 & 3.58$^{+0.14}_{-0.19}$ & $10.1745\pm0.0003$ & 5022$^{+140}_{-100}$ & $7.35\pm0.14$ & $1.40\pm0.1$ & $3.58\pm0.13$ \\
  211351097 & 132.478868 & 10.834468 & 1.1$^{+0.03}_{-0.16}$ & $12.2796\pm0.0002$ & 5949$^{+504}_{-60}$ & $1.37\pm0.03$ & $1.07\pm0.007$ & $1.15\pm0.07$ \\
  211351543 & 129.333431 & 10.842741 & 1.04$^{+0.04}_{-0.18}$ & $11.2813\pm0.0006$ & 6420$^{+640}_{-130}$ & $1.65\pm0.03$ & $1.11\pm0.008$ & $1.13\pm0.07$ \\
  \label{tab:tab1}
\enddata
\tablenotetext{1}{\tbf{Masses and radii from method described in Sec. \ref{sec:stellarparams}}}
\end{deluxetable*}

\subsection{Period Constraints and Priors} \label{sec:periodconstraints}
Due to the 960 day gap between C5 and C16 and the 104 day gap between C16 and C18, several transits may have been missed, although the exact number of missed transits is unknown. As a result, the period could be any \tbf{integer} division of the time difference between the two observed transits: 
\begin{equation}
\label{eq:1}
P = \frac{t_2 - t_1}{i}, \quad
P_{min} < P \leq t_2 - t_1
\end{equation} where $t_1$ is the mid-transit time of the first transit, $t_2$ is the mid-transit time of the second transit, $i$ is an integer and $P_{min}$ is the \tbf{minimum possible period, estimated by the} maximum length of observations before or after a transit in the data.

As a result, the period distributions of double transit planet candidates are highly multimodal, with peaks corresponding to each allowed integer $i$ in Eq. 1. \tbf{Furthermore, there are certain factors which skew the period distribution.} For instance, shorter period planets are favored because they are more likely to transit in the observation baseline. \citet{vanderburg2016} uses a period prior of the form $(B+D)/{P}$ for periods longer than the baseline in the case of a single transit with transit duration $D$ in a single baseline $B$. For the case of two observational baselines with exactly one transit in each baseline this is complicated further and is shown in \citet{becker2018}.

However, we have multiple planet systems, some which have data for three campaigns, C5, C16 and C18. As such, we fully generalized the \citet{becker2018} period prior to $n$ baselines, each either containing or not containing a transit. We find that the prior probability for a given orbital period can be expressed as follows:
$$Prob_{i}(P_{i}, P_{i,min}, D_{i}, B_{1}, B_{2}, ... , B_{n}, T_{i,1}, T_{i,2}, ... T_{i,n}) =  \displaystyle \prod_{j=1}^{n} A_j$$
$P_i$ = proposed period of planet i\\
$P_{i,min}$ = minimum period of planet i\\
$D_i$ = transit duration of planet i\\
$B_j$ = length of baseline j
\begin{displaymath}
T_{i,j} = 
\begin{cases}
0, & \text{if no transit of planet i exists in $B_j$}.\\
1, & \text{else}. \\
\end{cases} \text{,}
\end{displaymath}
\medmuskip=1mu
\thinmuskip=1mu
\thickmuskip=1mu
\small{\begin{equation}
  A_j=\begin{cases}
	\displaystyle 0, & \text{if $P_i \leq P_{i,min}$}.\\
    \displaystyle 1, & \text{if $P_i  > P_{i,min}$ and $P_i - D_i \leq B_j$}.\\
    
    \displaystyle 1-\frac{B_j+D_i}{P_i}, & \text{if $P_i  > P_{i,min}$, $P_i - D_i > B_j$, and $T_{i,j}=0$}.\\
    
    \displaystyle \frac{B_j+D_i}{P_i}, & \text{if $P_i  > P_{i,min}$, $P_i - D_i > B_j$, and $T_{i,j}=1$}.\\
  \end{cases}
\end{equation}}

\begin{figure}[h]
    \centering
    \plotone{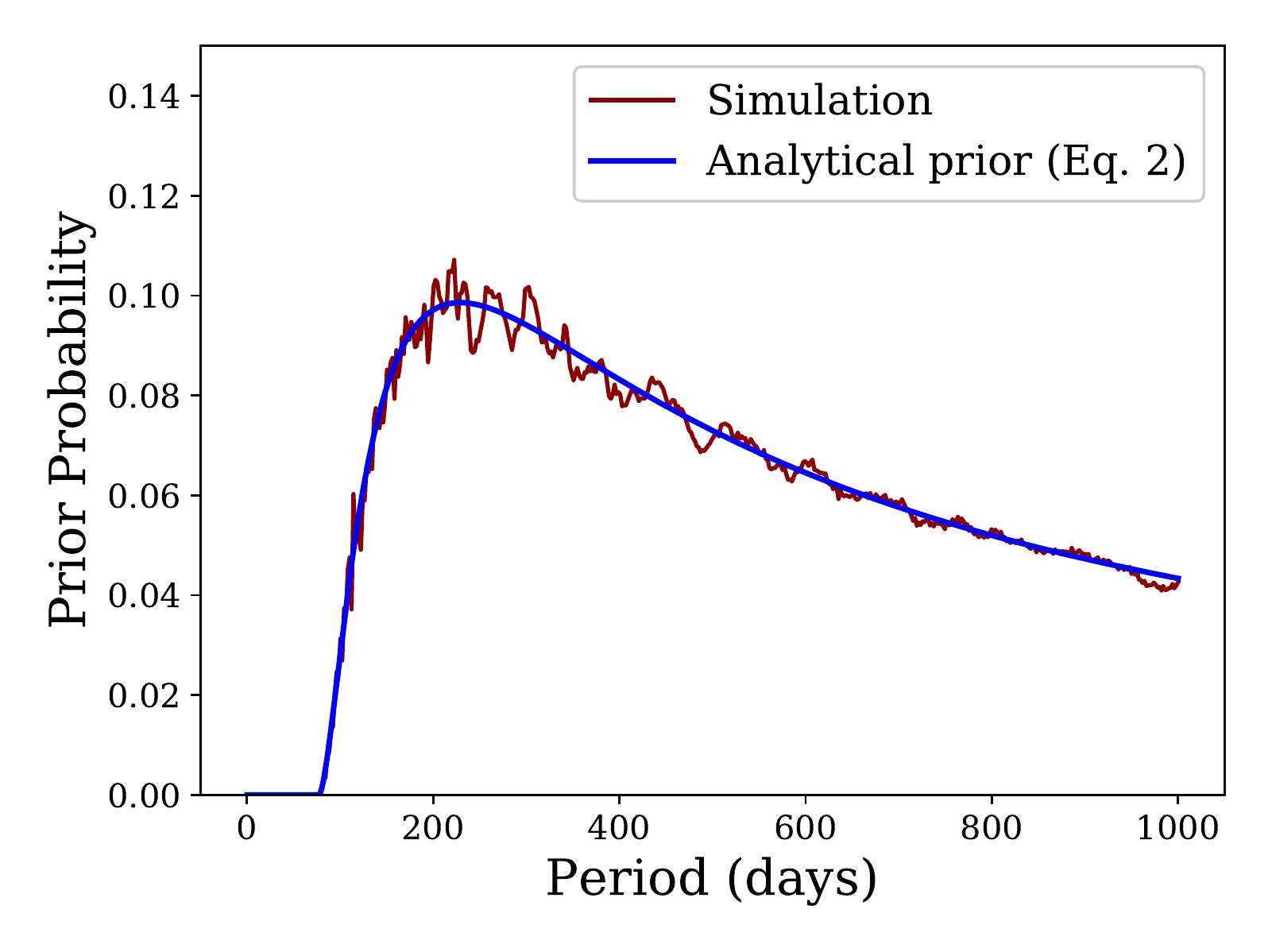}
    \caption{Simulation comparing orbital period prior probability and analytical prior probability presented in Section ~\ref{sec:periodconstraints}. This simulation is \tbf{based on three baselines with lengths equal to that of C5, C16, and C18; it also assumes one transit each in C5 and C16 and no transit in C18.}}
    \label{fig:periodprior}
\end{figure}

In order to validate this prior we performed a Monte Carlo simulation of the transit probabilities. We randomly drew 2000 periods in the interval (0.1,1000) days. For each of these periods, we drew 100 baseline separations in the interval [0,3000) days. We specified an observed set of transits ($T_j$ in Eq. 2), and tested the fraction of randomly drawn epochs which resulted in this configuration of transits. We show in Fig.~\ref{fig:periodprior} the results of this simulation for \tbf{baseline lengths the same as C5, C16, and C18 with one transit in C5 and C16 each but no transit in C18}.

\subsection{Stellar Parameters} \label{sec:stellarparams}
Using photometric and parallax data from Gaia Data Release 2 \tbf{(DR2; \citealt{gaiacollab2018})}, we \tbf{estimate} mass and radius values for each of the 6 stars. \tbf{We start with luminosities of each of the stars presented by \citealt{gaiacollab2018})}, and use empirical mass-luminosity relations presented in Eq. 4 of \citet{gafeira2012} to arrive at masses for each of the stars. We then find B and V magnitudes from two sources, \tbf{\citet{sampedro2017, nascimbeni2016}}, and obtain K magnitudes from 2MASS \citep{skrutskie2006}. We then find B-V and V-K color magnitudes and use the empirical relations provided in \citet{boyajian2014} to compute radii for both B-V and V-K magnitudes for each of the 6 stars. We find that these radii all are consistent with the DR2 reported radii, so we adopt \tbf{these} radii for all 6 stars, \tbf{due to the greater accuracy} of Gaia color-band photometry compared to the other photometry. We also use the B-V and V-K color photometry and empirical relations in \citet{boyajian2013} to compute effective temperatures for each of the 6 stars. Again, we find all values consistent with the Gaia reported value for \tbf{$T_{eff}$}, so we adopt the Gaia value. We note that one of the stars, EPIC211633458, appears to be an evolved star with a radius of $3.58\pm0.19 R_{\odot}$.

\subsection{Transit Modeling}
\label{sec:transitmodeling}
We use the BATMAN transit fitting Python package \citep{kreidberg2015} \tbf{based on the transit model of} \citet{mandel2002} in order to model our transit photometry. Our model assumes seven parameters per planet: transit time, orbital period, planet radius relative to stellar radius ($R_p/R_*$), transit duration, and impact parameter (the last two of which are later reparameterized as semi-major axis scaled to the stellar radius ($a/R_*$) and inclination). Additionally, we include four global parameters per system: a baseline offset parameter, a noise parameter (in order to encapsulate systematic effects in our photometric uncertainties), and two quadratic limb darkening parameters, using the parameterization presented in \citet{kipping2013b}. We jointly fit all the campaign lightcurves for each system using the period prior described in Section ~\ref{sec:periodconstraints}. \tbf{We also apply a Gaussian prior on stellar density by comparing the density estimated from the transit model to the spectroscopically determined density and uncertainties (see Section ~\ref{sec:stellarparams}). Further, for eccentricity we adopt a beta prior as described in \citet{kipping2013a}. Lastly, in the case of the candidate multiplanet system EPIC211939692, we assert that any system architecture with Hill radius crossings is impossible. In order to determine Hill radii, we used \citet{wolfgang2016} to estimate planet masses from planet radii.}

We explore our parameter space using MultiNest \citep{multinest2009} to sample the posteriors, as MultiNest is optimized for handling multimodal distributions such as our period distributions. In detail, we run MultiNest with constant efficiency mode, importance nested sampling mode, and multimodal mode all set to true, with a sampling efficiency of 0.01, 1000 live points, and an evidence tolerance of 0.1.

\subsection{Single \tbf{Planet Systems}}
\label{sec:singleplanets}
We find 5 systems with a single transiting planet that transits twice, one transit in each of two different campaigns. EPIC 211953574 was the only star for which we had three baselines of observation: C5, C16 and C18. For this system, we find matching transits in C5 and C16, but we observe no transits in the C18 baseline. Despite no transits occurring in C18, we can use the lack of a transit in C18 to constrain the period space further than would have been possible with only C5 and C16 data. The period distribution for this planet favors longer periods than the other four distributions. We find the planet to have a median period of 84.374 days and a radius of $2.74\pm{0.4}$ $R_e$ from our fits presented in Section ~\ref{sec:transitmodeling}.

We observe transits for EPIC211821192.01 in both C5 and C18. We find it to have a median period of 54.426 days, and a radius of $2.95\pm0.4$ $R_e$. EPIC211821192 is very similar to the Sun, with radius of $0.98^{+0.03}_{-0.01}$ R$_\odot$ and luminosity of $0.97\pm0.02$ $L_\odot$.

EPIC211633458.01 has a radius of $11.83\pm0.4$ $R_e$, making it the only giant candidate on our list. Because EPIC211633458 is an evolved star, despite a similar transit depth to the other candidates, the planet's radius is similar to that of Jupiter.

EPIC211351097 and EPIC 211351543 both present \tbf{a single transit in C5 as well as C18}.
We report comparisons for all five single planet systems between our transit parameter values and the values presented in \citet{lacourse2018} in Table ~\ref{tab:tab3}.

\tbf{As all the planet candidates in these systems have uncertain period estimates, we cannot calculate for certain whether these planet candidates would be in the habitable zones of their host stars. However, using the method used by the NASA Exoplanet Archive (https://exoplanetarchive.ipac.caltech.edu), derived from \citet{kasting1993}'s estimate of the habitable zones for these host stars in AU, we can place a probability on this, using Kepler's third law to convert our period distributions into distributions for semi-major axis. We present these probabilities in Table \ref{tab:tab2}.}

\begin{deluxetable}{ccccc}
\tablecolumns{5}
\tablewidth{0pt}
\tablecaption{Transit Parameters for Single Planet Systems}
\tablehead{
\colhead{} & \multicolumn{2}{c}{\citet{lacourse2018}} & \multicolumn{2}{c}{This work} \\
\colhead{EPIC ID} & \colhead{Epoch} & \colhead{Depth} & \colhead{Epoch} & \colhead{Depth} \\
 & BJD & ppm & BJD & ppm
}
\tablecolumns{5}
\startdata
  211953574 & 2457193.5608 & 590 & 2457193.588  & 400 \\
  211821192 & 2457177.1743 & 1103 & 2457177.203 & 784 \\
  211633458 & 2457201.6513 & 1197 & 2457201.776  & 900 \\
  211351097 & 2457161.8308 & 905 & 2457161.856 & 625 \\
  211351543 & 2457206.7598 & 491 & 2457206.776 & 400
  \label{tab:tab3}
\enddata
\end{deluxetable}

\subsection{EPIC 211939692, a Four Planet System}
One system of particular interest is the candidate system EPIC 211939692. \citet{lacourse2018} reported the system as an ``apparent multi-planet system" but did not specify further. We find that there are 6 transits in C5, and we find 3 additional transits in C18. 

\begin{figure*}[h]
    \centering
    \plotone{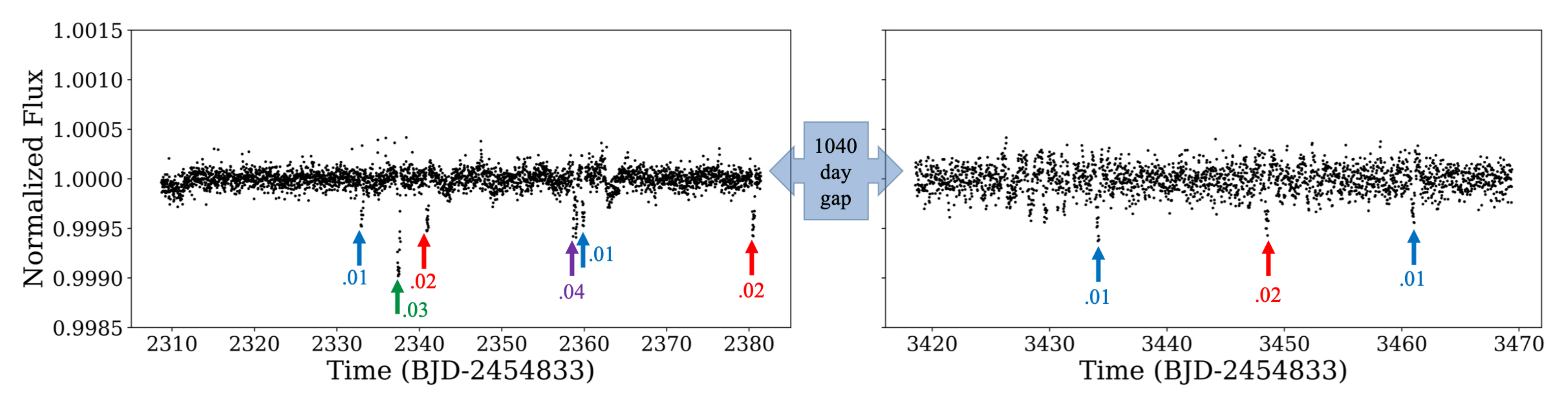}
    \caption{Light curve of EPIC 211939692 with transits labeled.}
    \label{fig:lightcurve}
\end{figure*}

We found that the first and fifth transits of the C5 lightcurve for EPIC 211939692 had \tbf{(by eye)} a similar transit duration and depth. The two transits are spaced apart by \tbf{$\sim$}26.854 days, and \tbf{extrapolating from this period} we find that the first and third transits in C18 \tbf{align with} the same ephemeris and also have similar depths and durations. 

In addition, the second and sixth transits in C5 are also similar \tbf{in depth and duration} with time difference \tbf{$\sim$}39.554 days; extending this forwards in time we find that the second transit in C18 fits this period and has the same depth and duration.

\begin{figure*}[h]
    \centering
    \plotone{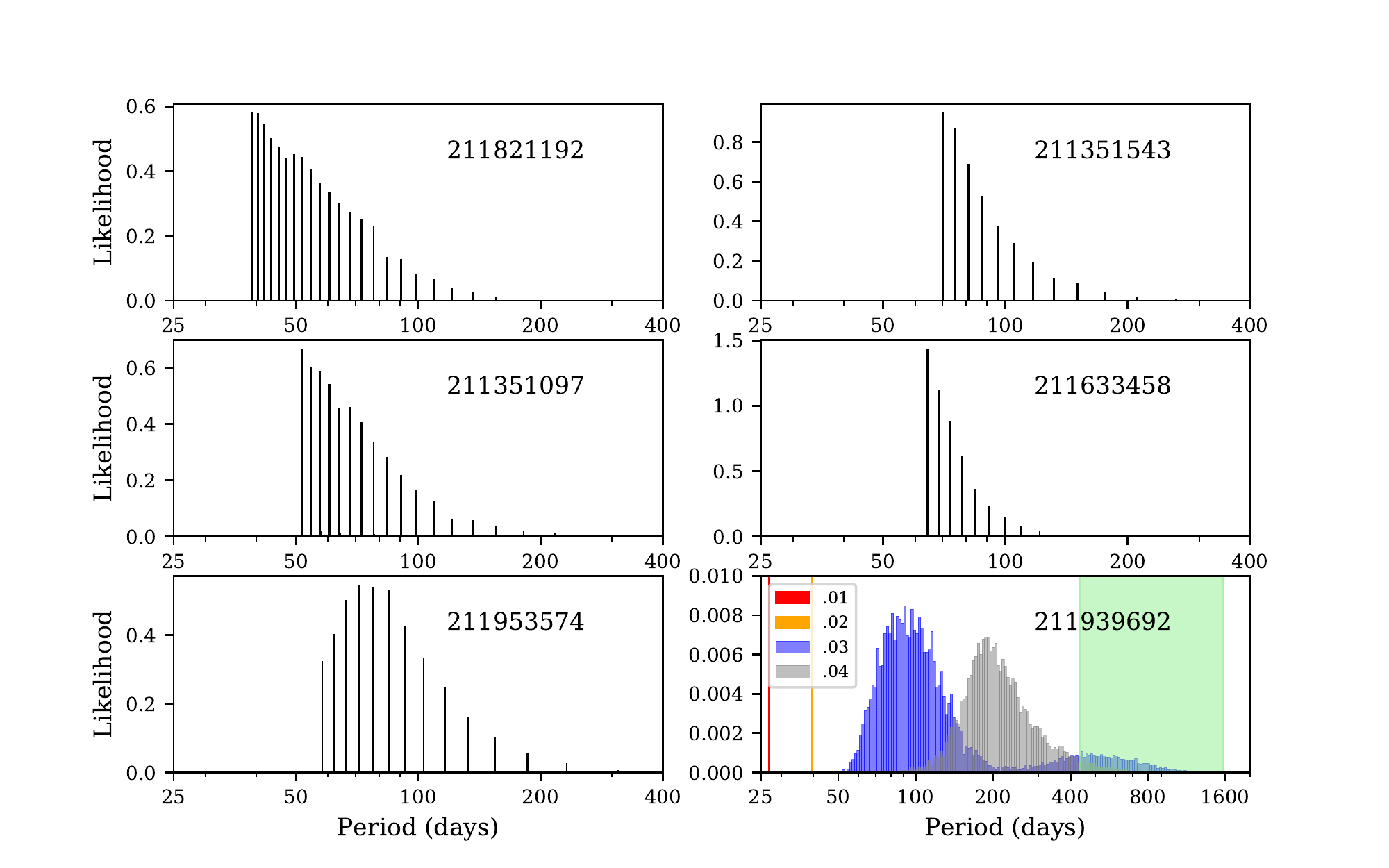}
    \caption{Period distributions for all planet candidates reported in this paper. \tbf{In the bottom right panel, the habitable zone is plotted in green to demonstrate that a portion of the period posterior distribution for the outer two planets is in the habitable zone (specifically, 37\% for the 211939692.03 and 12\% for 211939692.04).}}
    \label{fig:periods}
\end{figure*}

\tbf{The two outer planets in this system also have uncertain period estimates as they transit only once in the available data; we use the method described above in Sec. \ref{sec:singleplanets} to find probabilities that these planet candidates would fall within the habitable zone of their host star. We find probabilities of 37\% and 12\% for the two outer planets. These results are summarized in Table~\ref{tab:tab2} and shown in Fig. \ref{fig:periods}.}

We successfully performed a fit using the procedure outlined in Section ~\ref{sec:periodconstraints} to the C5 and C18 lightcurve on these two planets and found planetary parameters shown in Table ~\ref{tab:tab2}. This leaves two planet candidates, EPIC 211939692.03 and EPIC 211939692.04, that transit a single time in C5 and do not exhibit transits in C18. It is possible to constrain the period of these two planets further using the duration of the transit to approximate the period and using the fact that we do not observe a transit in the C18 observations. This is beyond the scope of this paper. We plan to return to the EPIC 211939692 system in an upcoming publication.

\newpage
\begin{deluxetable*}{cccccccccc}
\rotate
\tabletypesize{\tiny}
\tablecolumns{10}
\tablewidth{0pt}
\tablecaption{Planetary Parameters}
\tablehead{
\colhead{EPIC ID} & \colhead{211939692.01} & \colhead{211939692.02} & \colhead{211939692.03} & \colhead{211939692.04} & \colhead{211953574.01} & \colhead{211821192.01} & \colhead{211633458.01} & \colhead{211351097.01} & \colhead{211351543.01}
}
\startdata
\bf{Median Period} & 26.8549$\pm$0.0002 & 39.5530$\pm$0.0003 & 231.617 & 292.555 & 84.374  & 54.426 & 72.881 & 68.029 & 80.941 \\
\bf{Period Range} & - & - & 88.657-1101.691$^1$ & 167.635-651.952$^1$ & 58.007-185.626$^{1}$ & 47.327-108.856$^{1}$ & 64.307-109.324$^{1}$ & 51.831-155.502$^{1}$ & 70.148-175.370$^{1}$ \\
\bf{Epoch} & 2457166.040$\pm$0.007 & 2457173.980$\pm$0.005 & $2457170.483\pm0.004$ & $2457191.892\pm0.007$ & 2457193.588$\pm$0.008 & 2457177.203$\pm$0.004 & 2457201.776$\pm$0.01 & 2457161.856$\pm$0.03 & 2457206.776$\pm$0.005 \\
\bf{$e$} & 0.16$\pm$0.1 & 0.06$\pm$0.1 & - & - & 0.2$\pm$0.2 & 0.1$\pm$0.1 & 0.1$\pm$0.3 & 0.2$\pm$0.2 & 0.2$\pm$0.2 \\
\bf{$b$} & 0.18$\pm$0.1 & 0.23$\pm$0.1 & - & - & 0.80$\pm$0.2 & 0.45$\pm$0.20 & 0.88$\pm$0.06 & 0.54$\pm$0.2 & 0.80$\pm$0.2 \\
\bf{$\omega$} & $270^{+27}_{-30}$ & $197^{+22}_{-70}$ & - & - & $121^{+54}_{-39}$ & $139^{+73}_{-48}$ & $106^{+38}_{-30}$ & $103^{+71}_{-52}$ & $128^{+61}_{-44}$ \\
\bf{$R_p$/$R_s$} & 0.018$\pm$0.0007 & $0.021\pm{0.0007}$ & - & - & 0.020$\pm$0.002 &  0.028$\pm$0.002 & 0.030$\pm$0.003 & 0.025$\pm$0.006 & 0.020$\pm$0.002 \\
\bf{$R_p$ ($R_\oplus$)} & 2.60$\pm$0.09 & $2.94\pm0.10$ & $4.44\pm0.2$ & $2.62\pm0.1$ & 2.74$\pm$0.4 & 2.95$\pm$0.2 & 11.83$\pm$1.1 & 2.98$\pm$0.7 & 2.27$\pm$0.2 \\
\bf{$T_{eq}$(a=0.3) (K)} & 750 & 650 & 412 & 365 & 486 & 500 & 722 & 505 & 484 \\
\bf{$T_{eq}$(a=0.5) (K)} & 670 & 598 & 379 & 335 & 446 & 459 & 664 & 465 & 445 \\
\bf{$P_{HZ}$} & $<0.01$ & $<0.01$ & 37\% & 12\% & 0.8\% & $<0.01$\% & $<0.01$\% & 0.4\% & 0.4\% \\
\bf{$a/R_*$} & 30.6$\pm$0.8 & 40.7$\pm$0.9 & 101.5 & 129.4 & 62.2 & 56.1 & 20.2 & 58.0 & 73.5 \\
\bf{$a/R_*$ Range} & - & - & 56.9-357.0$^1$ & 89.2-250.0$^1$ & 21.2-140.0$^1$ & 42.9-93.6$^1$ & 8.0-30.1$^1$ & 26.5-116.0$^1$ & 35.5-139.4$^1$ \\
  \label{tab:tab2}
\enddata
\tablenotetext{1}{\normalsize 95\% confidence intervals. We report confidence intervals for planets which do not have uniquely-constrained periods and semi-major axes (see Fig. ~\ref{fig:periods})}
\end{deluxetable*}

\subsection{Verification of Transit Fitting}
\tbf{In order to verify the accuracy of the transit fitting procedure and period prior, we simulated double transit candidates using long period planets from the {\em Kepler} Mission. We used Kepler 420b and Kepler 538b, reported in \citet{santerne2014} and \citet{mayo2019} respectively. These planets have published periods of 86.6 and 81.7 days respectively. We clipped the light curves approximately to the baselines of C5 and C18 and the separation between the two campaigns. We then applied the fitting procedure to verify that the modes we detect contain the published value of period for these systems. In both cases the correct period is a mode on the period distribution as expected, verifying the results of out fitting procedure. We find the probability of the true period mode for K-420Ab = 8.4\% and for K-538b = 5.1\%. We note that we do not expect a majority of the distribution to be at the true mode because of the large number of modes.  We also confirm a high eccentricity (e = 0.77) for Kepler 420b as reported in \citet{santerne2014}.}
\section{Summary and Discussion} \label{sec:summary}

The fields of view of the K2 mission Campaigns 5, 16, and 18 covered overlapping sky area. We took advantage of this overlap to search for additional transits of the single-transit candidates reported by \citet{lacourse2018} from Campaign 5. Six systems displayed additional transits, one of which is a multiplanet system with 4 candidate planets. We present a method to constrain the period of transiting objects in gapped photometric datasets, which we apply to 5 systems. We present these period constraints in Section ~\ref{sec:periodconstraints}. 

These six systems are amongst the longest period planetary systems discovered by the K2 mission to date. Only EPIC248847494b \citep{giles2018} and potentially K2-263b \citep{mortier2018} and K2-118b \citep{dressing2017} have longer periods. Of the 9 planet candidates found, 4 have constrained minimum periods (see Eq. \eqref{eq:1}) longer than K2-263b and K2-118b. The planets reported on in this work occupy a sparse region of parameter space, representing transiting long period planets orbiting stars which are moderately bright. This region of parameter space is particularly important because it contains the habitable zones of relatively bright sun-like stars with high propensity for characterization. One system of note is EPIC211633458: a relatively bright (V=10.38, K=8.27) host star to a giant planet. It has high potential for future follow up observations including radial velocity confirmation. Furthermore, its large radius, likely gaseous composition ($R_p = 2.98$), and brightness make future atmospheric characterization practicable.

\tbf{We also find a very interesting candidate multiplanet system around the star EPIC 211239692.
We can fully constrain the periods of two planet candidates (as they transit multiple times per campaign), as well as place broad constraints on the other two candidates. This system has high follow up potential with radial velocity observations  constrain the masses and find the periods of the two single-transiting planets, especially given their non-negligible probabilities of residing in the habitable zone.}

Radial velocity observations \tbf{of these systems could} allow an estimation of not only planet mass, but also potentially a unique constraint on the period. \tbf{However, whether or not these systems are amenable to followup may depend on whether the orbital period is short enough for the radial velocity semi-amplitude to be detectable.} How best to pursue radial velocity and other ground-based followup observations in order to uniquely constrain the period for systems such as these remains an open question, one we would like to address in a future publication.

These planets would also add to the short list of transiting long period exoplanet candidates. Such systems are useful to study theories of planet formation, particularly pertaining to the ``photoevaporation valley" of planetary radii near 1.6 to 2.0 Earth radii \citep{fulton2017}.

We note that the challenge of period constraints for these six \tbf{systems} in the gapped K2 dataset (see Section ~\ref{sec:periodconstraints}) is \tbf{a} highly pertinent problem for {\em TESS} observations because of the similar observational strategy. This procedure may be used as a template for future systems \tbf{of planets with multiple transits with intervening gaps in observation} to constrain the period. Because {\em TESS} observes \tbf{regions} of sky repeatedly, even observations with no transits can be used to constrain the period space. The EPIC 211953574 system, which has transits in C5 and C16 but not in C18, is a prime example of this technique. Only small areas of the sky can be continuously observed by current transit surveys sensitive to small planets, so it is \tbf{important} to make use of observations in these gapped datasets in order to maximize the planetary yield and add to the number of longer period planetary systems with transit-based constraints.

\newpage
\acknowledgments
We thank Andrew Vanderburg for helpful discussions and suggestions. A.W.M. is supported by the NSF Graduate Research Fellowship grant no. DGE 1752814. This research has made use of the NASA Exoplanet Archive, which is operated by the California Institute of Technology, under contract with the National Aeronautics and Space Administration under the Exoplanet Exploration Program. This work also made use of \texttt{lightkurve} \tbf{\citep{lightkurve}}, a software package for the reduction and analysis of {\em Kepler}, K2, and {\em TESS} data. This open source software project is developed and distributed by the NASA Kepler Guest Observer Office. This research was supported in part by NASA through the K2 Guest Observer program. C.D. acknowledges support from the Hellman Faculty Fund.

%

\vspace{5mm}
\facilities{Kepler, Gaia, 2MASS, NASA Exoplanet Archive, ADS}


 \software{
 astropy \citep{astropy2018}, K2SC \citep{Aigrain2016}, lightkurve \citep{lightkurve}, BATMAN \citep{kreidberg2015}, MultiNest \citep{multinest2009}
          }

\newpage
\bibliography{references}



\end{document}